\documentclass[12pt,a4paper,twoside]{article}
\usepackage{epsfig}
\newlength{\figwidth}
\newlength{\figheight}
\def\gg{$\gamma \gamma$}

\def\wgg{$W_{\gamma \gamma}$}

\def\ggg{$\Gamma_{\gamma \gamma}$}
\def\gggsm{$\Gamma_{\gamma \gamma}^{SM}$}

\def\pgg{$\phi_{\gamma \gamma}$}
\def\pggsm{$\phi_{\gamma \gamma}^{SM}$}

\def\epem{$e^+ e^-$}

\def\z0{Z}

\title{
       \vspace{-1.5cm}
       \begin{flushright}
       \begin{tabular}{l}
       {\large CERN-TH/2002-165 }    \\[-3mm]
       {\large IFT - 28/2002}    \\[-3mm]
       {\large hep-ph/0207294 }  \\[-3mm]
       {\large July 2002}
       \end{tabular}
       \end{flushright}
       \vspace{1.5cm}
      \sc  Study of the Higgs-boson decays into 
       $W^+W^-$ and $\z0  \z0  $  at the Photon Collider
       }
 \author{Piotr Nie\.zurawski, Aleksander Filip \.Zarnecki \\
 {\small\it Institute of Experimental Physics, Warsaw University, 
    ul. Hoza 69, 00-681 Warsaw, Poland} \\[3mm]
 Maria Krawczyk \\
 {\small\it Institute of Theoretical Physics, Warsaw University, 
        ul. Hoza 69, 00-681 Warsaw, Poland} \\[-2mm]
 {\small and} \\[-2mm]
 {\small\it Theory Division, CERN, CH-1211 Geneva 23, Switzerland}
 } 

\date{}

\begin{document} 

\maketitle 

\vfill

\begin{abstract}
Production of the Standard Model Higgs-boson at the photon collider at
TESLA is studied for  Higgs-boson masses above 150 GeV. A simulation of 
signal and background processes takes into account realistic luminosity 
spectra and detector effects. In the considered mass range, large 
interference effects are expected in the $W^+ W^-$ decay channel. By 
reconstructing $W^+ W^-$ and $\z0  \z0  $ final states,
not only the $h \rightarrow \gamma \gamma$
partial width can be measured, but also the phase of the scattering amplitude.
 This opens a new window onto the precise
determination of the Higgs-boson couplings.
Models with heavy, fourth-generation fermions and with enlarged  Higgs sector 
(2HDM~(II)) are considered.
\end{abstract}

\vfill

\begin{flushleft} 
CERN-TH/2002-165 \\
IFT - 28/2002    \\
hep-ph/0207294  \\
July 2002
\end{flushleft}


\thispagestyle{empty}

\clearpage

%
%

\section{Introduction}
\label{sec:intro}

A photon collider has been  proposed as a natural
extension of the \epem\ linear collider project TESLA \cite{tdr_pc}.
The physics potential of a photon collider is very
rich and complementary to the physics program of \epem\
and hadron colliders.
It is the ideal place to study the properties of the Higgs-boson
and the electroweak symmetry breaking (EWSB).
If the Standard Model Higgs-boson mass is below $\sim$140 GeV, 
its properties can be measured in detail using the  
$h \rightarrow b \bar{b}$ decay channel.
However, if its mass exceeds 170 GeV the $b\bar{b}$ branching ratio
goes below 1\% and we have to consider other decay channels for a
precise measurement of the Higgs-boson.
In this paper the feasibility of measuring  Higgs-boson production
using the $W^+ W^-$ and $\z0 \z0$ decay channels will be considered.
Because of  the interference with other Standard Model contributions, 
the process turns out to be sensitive not only to the 
Higgs-boson partial width to \gg, \ggg, but also to 
the phase of the $\gamma \gamma \rightarrow h $ amplitude, \pgg.
A precise measurement of both \ggg\ and \pgg\ is important
for a  unique determination of the  Higgs-boson couplings
and can turn out to be crucial in  distinguishing 
between different models of ``new physics'',
see also \cite{ginzburg,cros_zz,gounaris,asakawa}.

Prospects for measuring Higgs-boson production at the
Photon Collider in the $W^+ W^-$ and $\z0 \z0$ decay channels 
has been previously considered 
in \cite{ginzburg,belanger,morris,boos} ($W^+ W^-$)
and \cite{cros_zz,gounaris,jikia,berger,dicus} ($\z0 \z0$).
%
%
Current analysis is the first one to include realistic 
luminosity spectra and detector simulation, 
which is especially important in an accurate measurement
of  such a subtle effect, as the phase of the amplitude.


\section{Vector boson production}
\label{sec:vecbos}

In the narrow-width approximation
%
the total cross section 
for resonant Higgs-boson production in \gg\ collisions,
with subsequent Higgs decay into a  pair of massive vector bosons $VV$
($W^+W^-$ or $\z0  \z0 $) can be written as
\begin{eqnarray}
\sigma_{\gamma \gamma \rightarrow h \rightarrow VV }
& = &
\left.
\frac{1}{{\cal L}_{\gamma \gamma}}
\frac{d{\cal L}_{\gamma \gamma}^{J_z = 0}}
{d W_{\gamma \gamma}}
\right|_{W_{\gamma \gamma} = M_h}
\cdot
\frac{4\pi^2 \; \Gamma_{\gamma \gamma}}{M_h^2}
\cdot
BR( h \rightarrow VV ),  \label{eq:nwa}
\end{eqnarray}
where the first term describes the differential \gg\ luminosity 
spectra for the projection of the angular momentum of 
the \gg\ system on the beam axis $J_z$, $J_z = 0$; 
$M_h$ and \ggg\ are the mass and  the two-photon width of the Higgs-boson;
and $BR( h \rightarrow VV )$ is its branching ratio to $VV$.
%
\begin{figure}[bp]
  \begin{center}
  \epsfig{figure=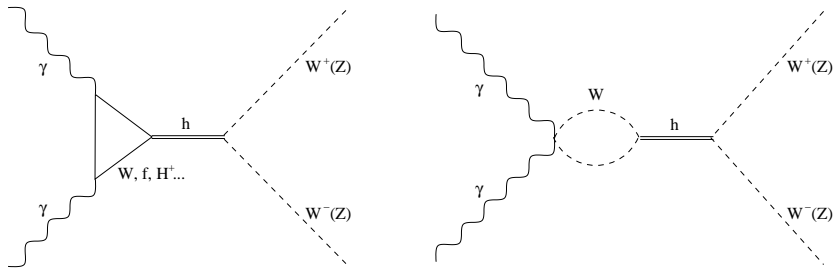,height=3.5cm,clip=}
  \end{center}
  \vspace{-0.5cm}
  \caption{Diagrams for $W^+W^-$  and $\z0  \z0 $ pair
production at the photon collider via the  production 
and decay of Higgs-bosons.}
  \label{fig:gghvv0}
\end{figure}
The Higgs-boson does not couple to \gg\ on the tree level.
The two-photon width \ggg\ describes the effective $\gamma \gamma h$
coupling resulting from the loop contributions of all charged particles,
as shown in Fig.~\ref{fig:gghvv0}. In the Standard Model (SM) the dominant
contribution comes from the $W^\pm$ and $t$ loops. However, if there are 
any new, heavy, charged particles, they will contribute to \ggg\ even
in the high mass limit.

\begin{figure}[p]
  \begin{center}
  \epsfig{figure=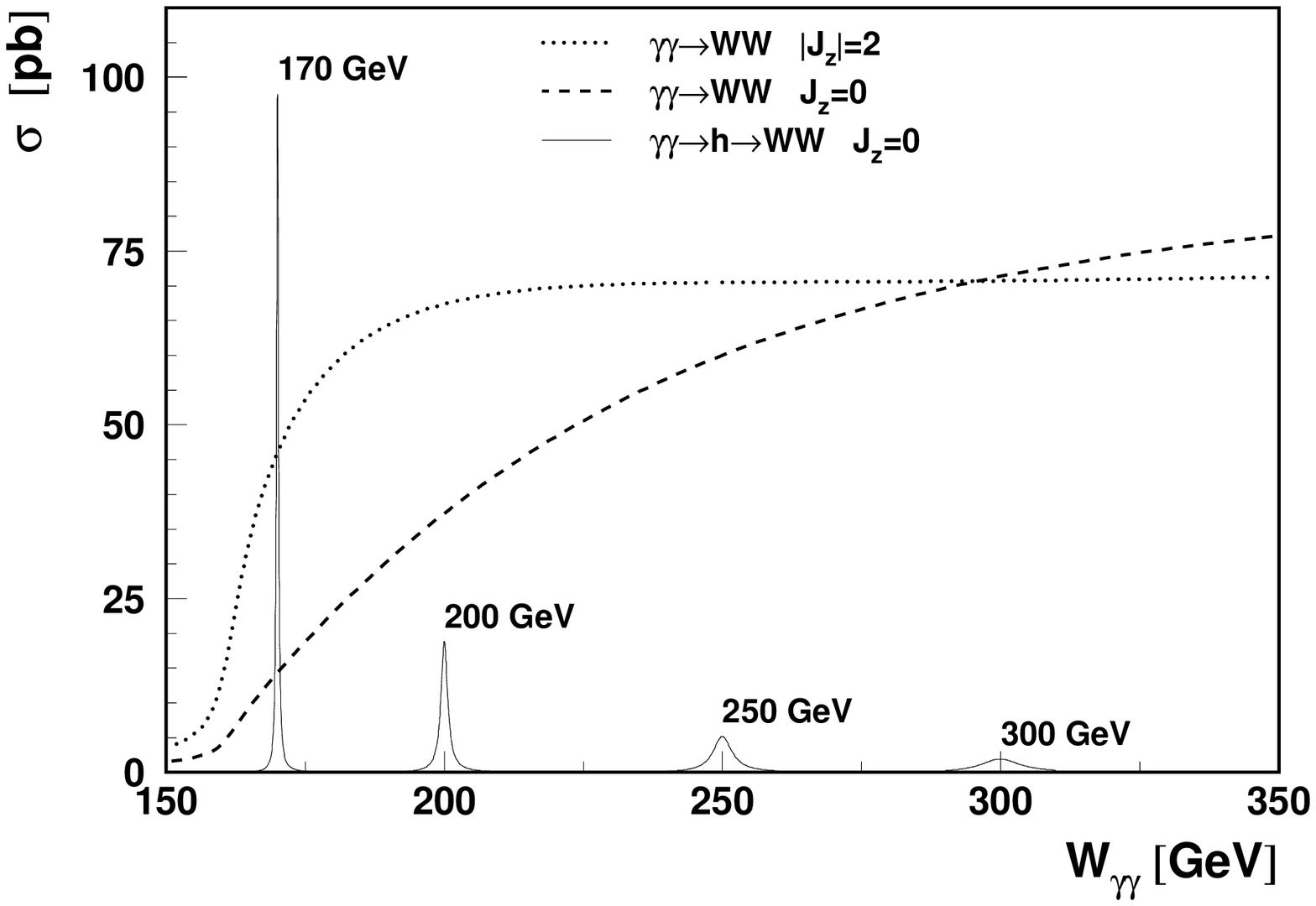,height=\figheight,clip=}

  \epsfig{figure=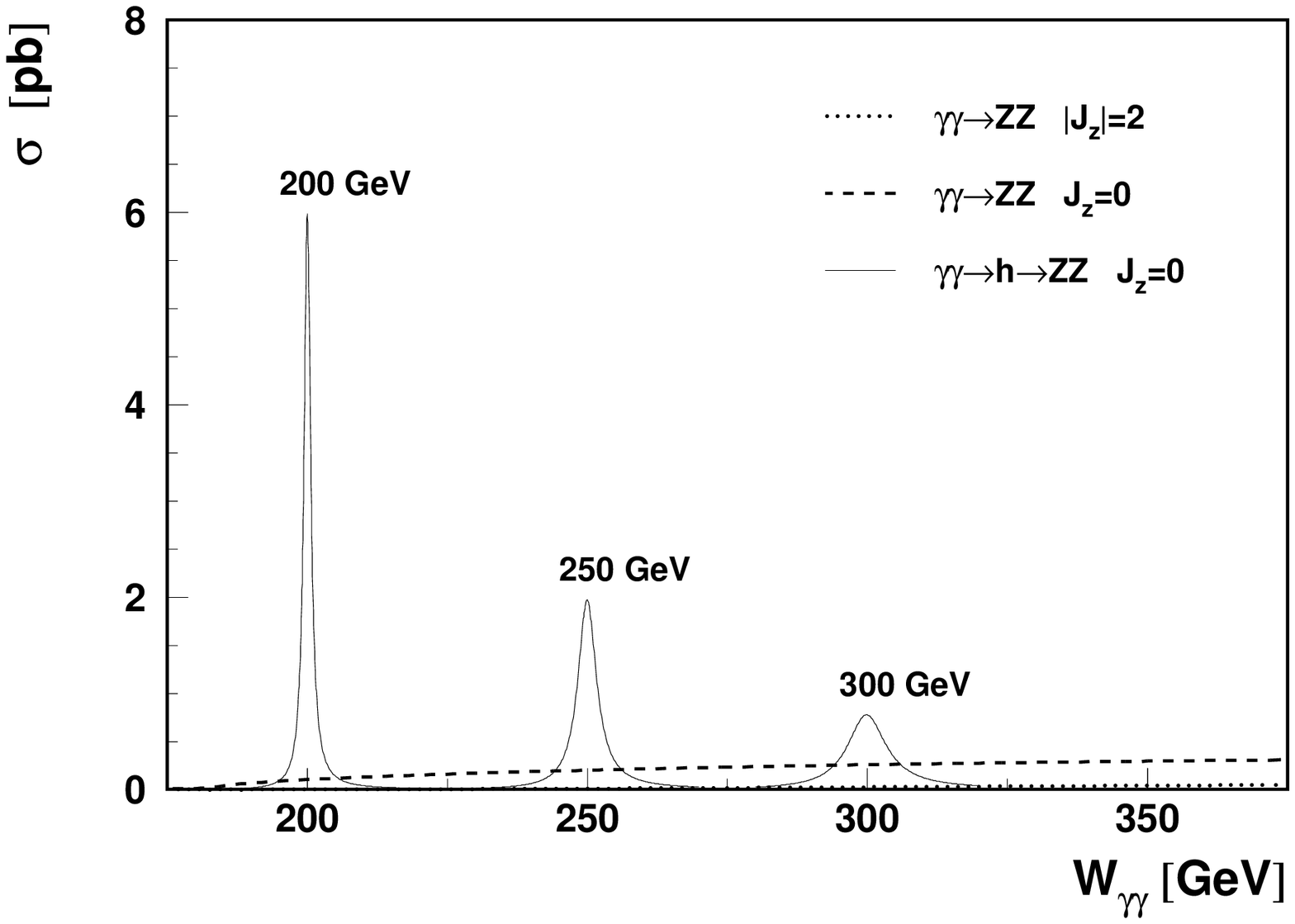,height=\figheight,clip=}
  \end{center}
  \caption{Cross section as a function of the photon--photon 
center-of-mass energy \wgg , for different processes contributing 
to the production of $W^+W^-$ pairs (upper plot) and $\z0  \z0 $ pairs 
(lower plot) at the photon collider.
Cross sections for direct vector boson pair-production, for total
angular momentum of colliding photons $J_z=0$ and $|J_z|=2$, are compared with
the cross section expected from the Higgs-boson production and decay,
for different Higgs-boson masses.
          }
  \label{fig:cros_wwzz1}
\end{figure}

Presented in Fig.~\ref{fig:cros_wwzz1} (solid line) are the cross sections
for the production of $W^+W^-$ (upper plot) and $\z0  \z0 $ 
(lower plot) pairs at the photon collider,
expected from the Higgs-boson production and decay,
as a function of the photon--photon center-of-mass energy \wgg.
Cross sections for  different Higgs-boson masses were evaluated using the  
HDECAY program \cite{hdecay} for calculating the SM Higgs-boson 
width and branching ratios.
The total width of the Higgs resonance
increases from about 0.4 GeV at $M_h = 170$ GeV to
about 8.5 GeV at  $M_h=300$ GeV.
%
%
However, the total cross section for Higgs-boson production in  
photon--photon collisions slowly changes with $M_h$.

\begin{figure}[tbp]
  \begin{center}
  \epsfig{figure=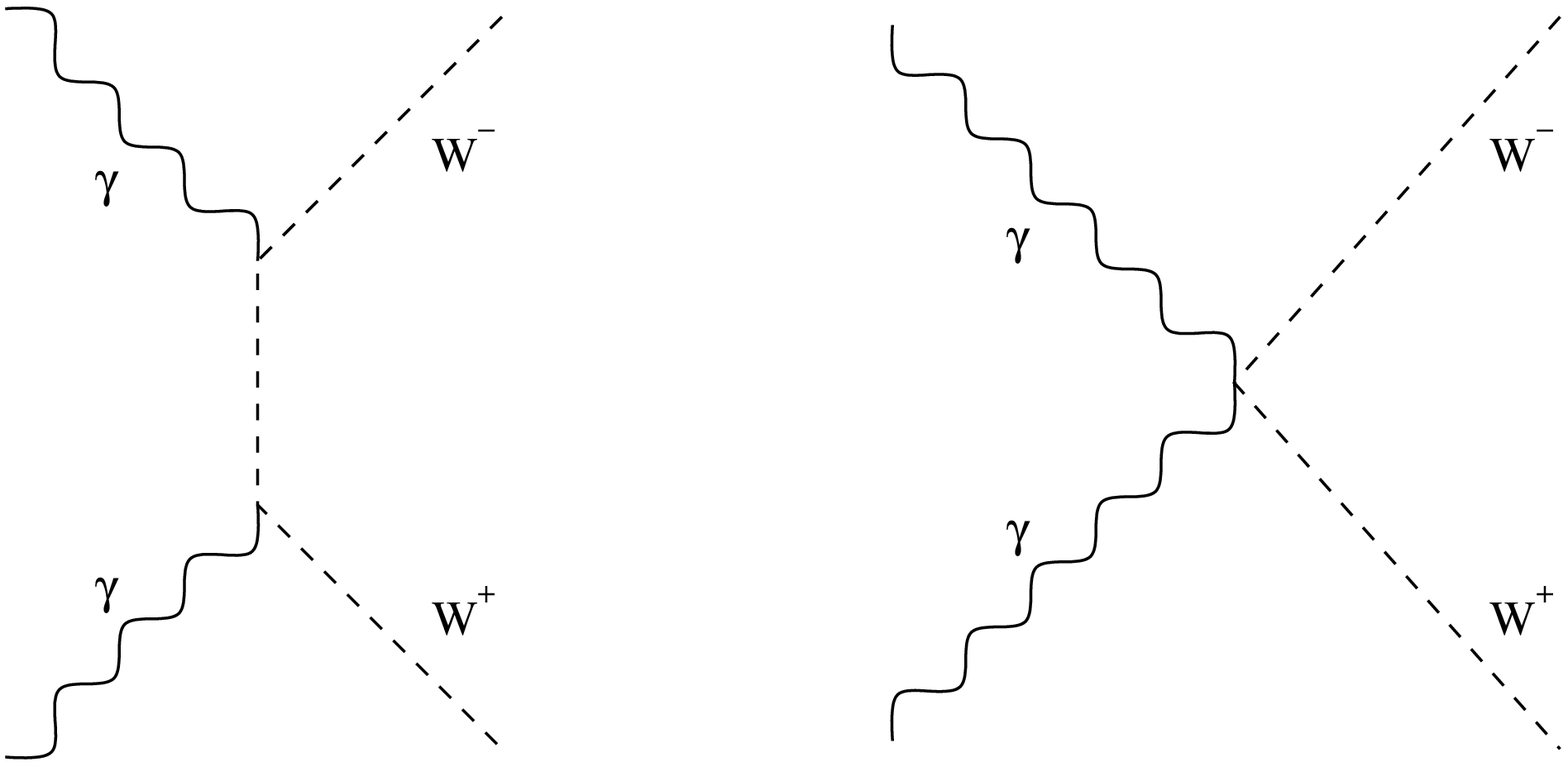,height=3.5cm,clip=}
  \end{center}
  \caption{Diagrams for direct $W^+W^-$ pair
production at the photon collider.}
  \label{fig:ggww}
\end{figure}

When studying vector-bosons production from the Higgs-boson decays, the
background from direct vector-bosons production has to be considered.
The non-resonant $W^+W^-$ pair production is a tree-level process and
is expected to be large.
The corresponding diagrams are shown in Fig.~\ref{fig:ggww}.
The cross sections for direct  $W^+W^-$  pair production 
for $J_z=0$ (dashed line) and $|J_z|=2$ (dotted line) are included 
in Fig.~\ref{fig:cros_wwzz1} (upper plot).
The calculation was
based on the scattering amplitudes presented in \cite{belanger,morris}.
%
For photons colliding with $|J_z|=2$ the cross section
for direct $W^+W^-$ pair production 
increases fast above the production threshold and then saturates
at the level of about $80\; \rm pb$.
The cross section for $J_z=0$ increases slowly with the energy and reaches
the same order of magnitude only at $W_{\gamma \gamma} \sim$ 300 GeV.
For Higgs-boson masses above about 190~GeV, the non-resonant cross section
exceeds the cross section for  $W^+ W^-$ pair production 
from Higgs-boson decay at  \wgg $=M_h$,
$\gamma \gamma \rightarrow h \rightarrow W^+ W^-$.

Direct $\z0  \z0 $ production is only possible at the loop level.
The corresponding cross sections, 
for $J_z=0$ (dashed line) and $|J_z|=2$ (dotted line), are included 
in Fig.~\ref{fig:cros_wwzz1} (lower plot).
The cross section for  $J_z=0$ is only of the order of $100 ~\rm fb$.
Therefore, the total cross section for $\z0  \z0 $ pair production
at $W_{\gamma \gamma} \sim M_h$ is dominated by the Higgs-boson contribution,
for all considered values of the Higgs-boson mass.
In the analysis presented here the asymptotic cross-section formula 
for  $\gamma \gamma \rightarrow \z0  \z0 $
is used \cite{cros_zz}, neglecting terms of 
${\cal O}(m_Z^2/W_{\gamma \gamma}^2)$

\begin{figure}[tbp]
  \begin{center}
  \epsfig{figure=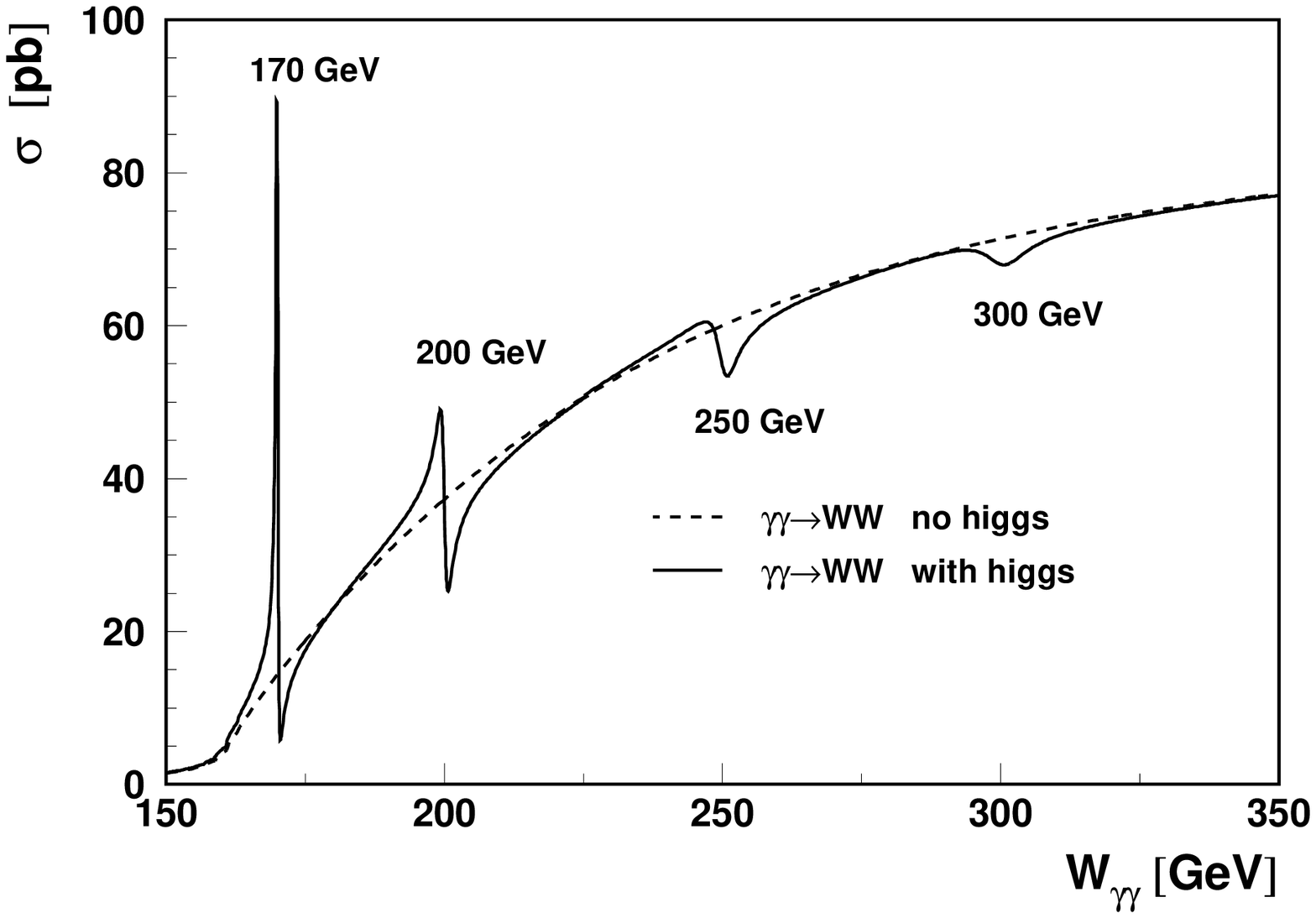,height=\figheight,clip=}

  \epsfig{figure=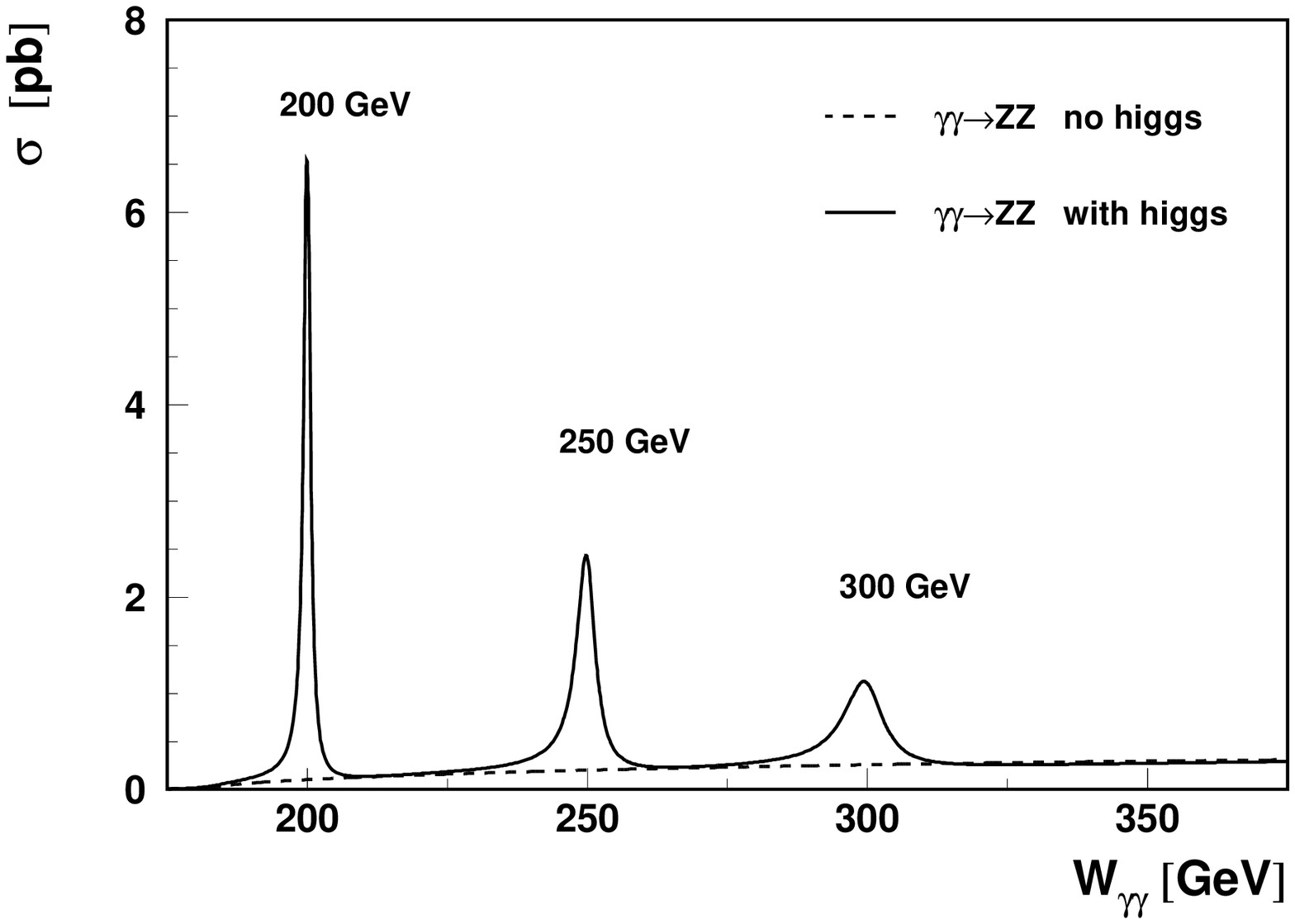,height=\figheight,clip=}
  \end{center}
  \caption{Total cross section as a function of the photon--photon 
center-of-mass energy \wgg , for $W^+W^-$ pair (upper plot) and 
$\z0  \z0 $ pair (lower plot) production at the photon collider,
for a total angular momentum of the colliding photons of $J_z=0$.
Cross section calculated for different Higgs-boson masses, as indicated
in the plots (solid line) is compared with the cross section 
expected from the direct vector boson pair-production (dashed line).
          }
  \label{fig:cros_wwzz2}
\end{figure}

For photons colliding with $J_z=0$ the interference between
the signal of vector-boson production via the Higgs resonance
and the background from direct vector-boson production has to be taken
into account.
Shown in Fig.~\ref{fig:cros_wwzz2} are the total cross sections
for $W^+W^-$ (upper plot) and $\z0  \z0 $ (lower plot) 
pair production at the photon collider,
for a  total angular momentum of colliding photons $J_z=0$,
 as a function of the photon--photon center-of-mass energy \wgg .
For  $W^+W^-$ pair production, very large interference effects 
are observed.
For Higgs-boson masses above about 200 GeV, the negative contribution
from the interference term is larger than the resonant (Higgs-boson only)
contribution and results in the decrease of the total 
$W^+W^-$ pair production cross section. 
For  $\z0  \z0 $ pair production the contribution of the interference 
term is also visible, resulting in an  asymmetric Higgs resonance,
but the contribution of the interference term to the total
cross section is small.
%

\begin{figure}[tbp]
  \begin{center}
  \epsfig{figure=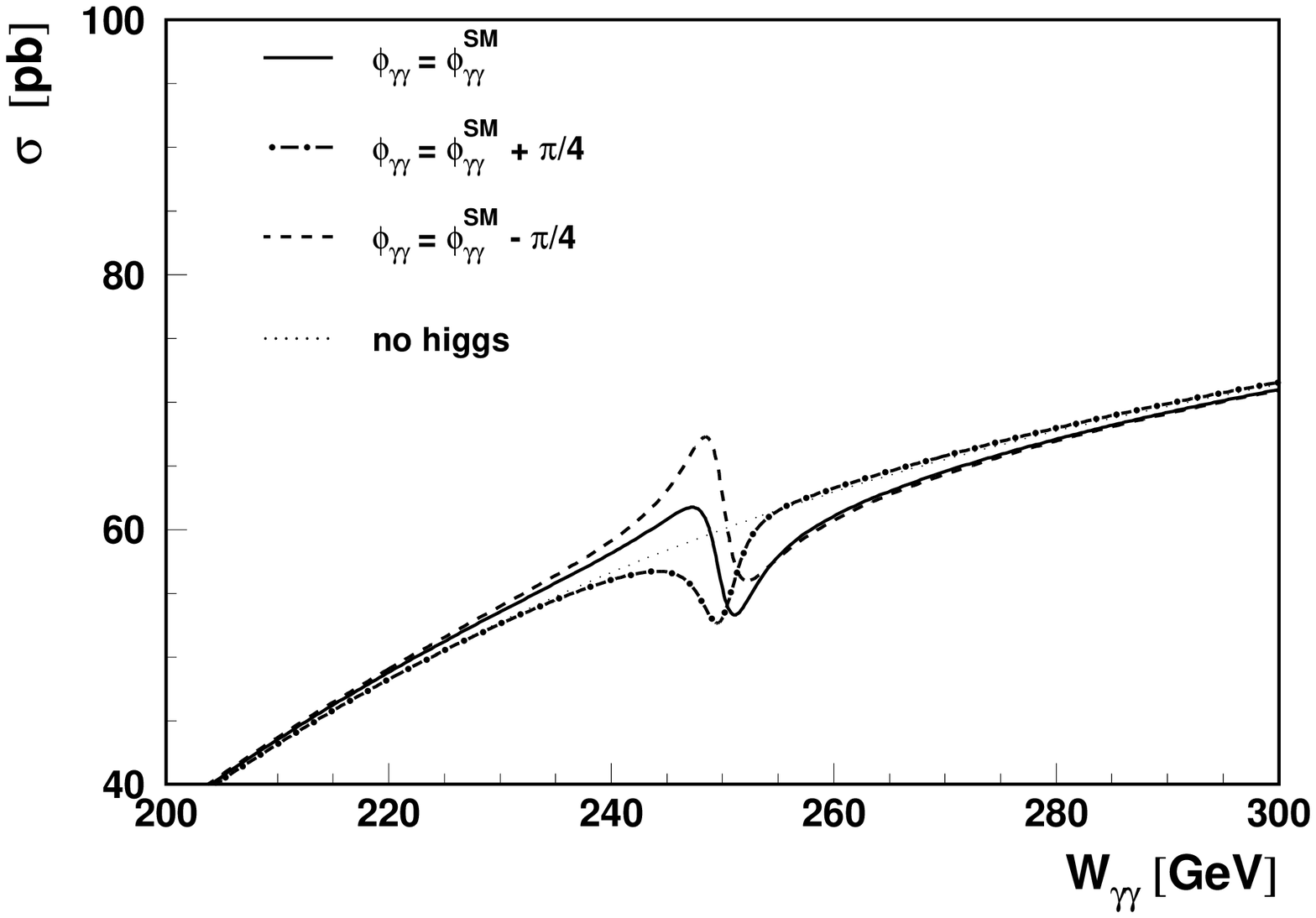,height=\figheight,clip=}
  \end{center}
  \vspace{-1cm}
  \caption{Total cross section as a function of the photon--photon 
center-of-mass energy \wgg , for $W^+W^-$ pair production at the 
photon collider, for $M_h =  250$ GeV and $J_z=0$.
The cross section is calculated for an  amplitude phase \pgg\ equal to the
Standard Model  prediction \pggsm\ 
and for a phase shifted by $\pm \pi / 4$.
The cross section for direct  $W^+W^-$ pair production (without Higgs-boson 
contribution) is included for comparison.
          }
  \label{fig:cros_wwph}
\end{figure}

The cross section for the Higgs-boson production in the \gg\ collisions
is proportional to the two-photon Higgs-boson  width \ggg.
From the measurement of the $W^+W^-$ and $\z0  \z0 $ pair production 
cross sections at the Higgs resonance, the value of 
\ggg\ can be extracted.
However, the measurement of the interference term contributions
allows us to access an additional piece of information about the phase of
the $h \rightarrow \gamma \gamma $ amplitude, \pgg.
The sensitivity of the total $W^+W^-$ pair production 
cross sections to the phase \pgg\ is illustrated in Fig.~\ref{fig:cros_wwph}.
Deviations of \pgg\ from the value predicted by the Standard Model can
significantly affect both the total $W^+W^-$ pair production cross section 
and the shape of the differential cross section, although the 
two-photon width \ggg\ is not changed.
The measurement of \pgg\ thus gives us more information about
the Higgs-boson couplings, complementary to the measurement of \ggg.
%

\begin{figure}[tbp]
  \begin{center}
  \epsfig{figure=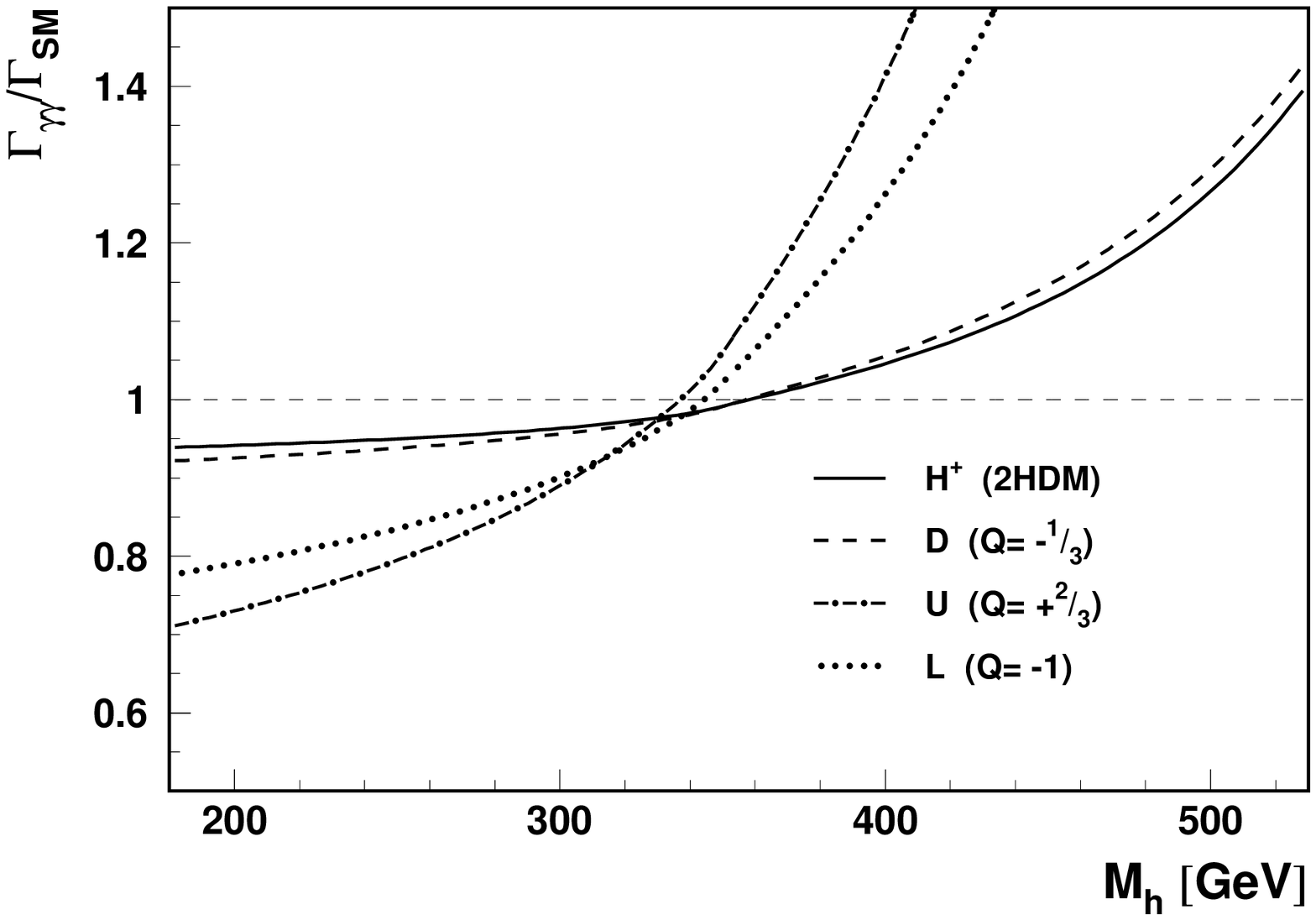,height=\figheight,clip=}
  \end{center}
  \caption{The ratio of the Higgs-boson decay width \ggg\
    to the width predicted in the Standard Model \gggsm\
    as a function of the Higgs-boson mass $M_{h}$, for
    models with additional heavy charged particles (fourth-generation
    $d$-quark $D$, $u$-quark $U$ or lepton $L$, or charged Higgs-boson of the
    Standard Model-like 2HDM~(II)).  }
  \label{fig:gam_exp}
\end{figure}

\begin{figure}[tbp]
  \begin{center}
  \epsfig{figure=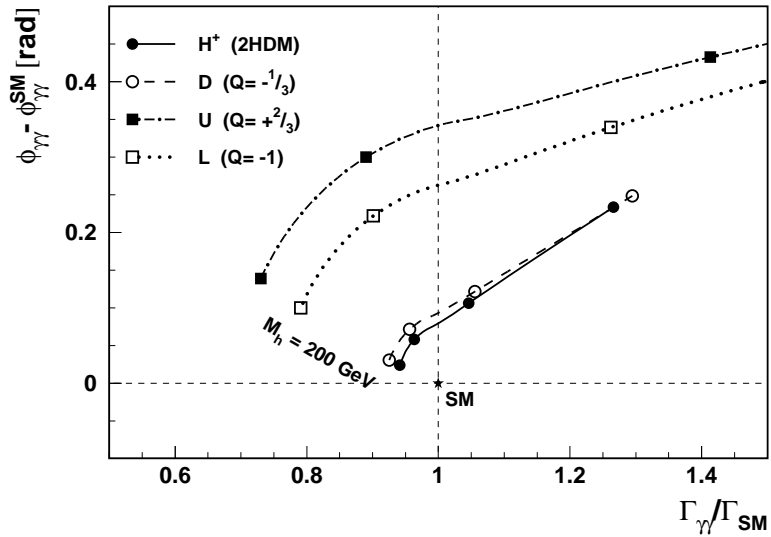,height=\figheight,clip=}
  \end{center}
\caption{The difference between 
\pgg\ and the phase predicted in 
 the Standard Model \pggsm\ as a function of the decay width 
 ratio \ggg/\gggsm. 
 Models with additional heavy charged particles (fourth-generation
$d$-quark $D$, $u$-quark $U$ or lepton $L$, or charged Higgs-boson of the
 Standard Model-like 2HDM~(II)) with mass of 800 GeV are considered.
 The curves were calculated
 by changing the Higgs-boson mass $M_{h}$ from 200 to 500~GeV; symbols
 are plotted every 100 GeV.
          }
  \label{fig:gam2phi_exp}
\end{figure}

Shown in Fig.~\ref{fig:gam_exp} is the ratio of the Higgs-boson decay 
width \ggg\ to the width predicted in 
the SM  with three fermion families,  \gggsm,
as a function of the Higgs-boson mass $M_{h}$, for
models with additional contribution due to one extra charged particles.
The new particles considered 
are heavy, fourth-generation fermions 
(fourth-generation up quark $U$, down quark $D$ or charged lepton $L$)
or the charged Higgs-boson $H^+$ of the Standard Model-like 
Two-Higgs-Doublet Model 2HDM~(II) (couplings of the lightest
Higgs-boson to fermions, $W^\pm$ and $Z$ are the same as in the
SM; see also \cite{2HDM}).
In all cases  the mass of the new particle 
(one new particle in each model)
is set to 800 GeV.
For Higgs-boson masses below about 330 GeV the contribution of the new heavy
charged particle to the  $h \rightarrow \gamma \gamma $ vertex
reduces the decay width \ggg,
whereas for masses above about 360 GeV \ggg\ increases in
all models.
For Higgs-boson masses around 350~GeV, the sensitivity
of \ggg\ to the new particles turns out to be significantly reduced,
as for all considered models predicted \ggg\ value 
is close to the SM expectations.
However, significant deviations from the SM predictions
are expected in this mass range 
for the  phase of the $h \rightarrow \gamma \gamma $ amplitude,
as shown in Fig.~\ref{fig:gam2phi_exp}.
If, for a Higgs-boson mass around 350 GeV, it happens that all couplings
of the Higgs-boson and the $h \rightarrow \gamma \gamma $ width
are the same as in the SM then  the measurement of
that phase is the only way to distinguish between different models.
%

In all cross-section calculations the electromagnetic coupling constant
at the Thomson limit was used, $\alpha_{em} \approx 1/137$,
%
as the relevant scale is the photon virtuality
$Q^2 = 0$. \footnote{We would like to thank M. Spira and J. Kalinowski
for  enlightening discussions on this point.}


\clearpage

\section{Analysis}
\label{sec:anal}

\subsection{Luminosity spectra}

\begin{figure}[tbp]
  \begin{center}
  \epsfig{figure=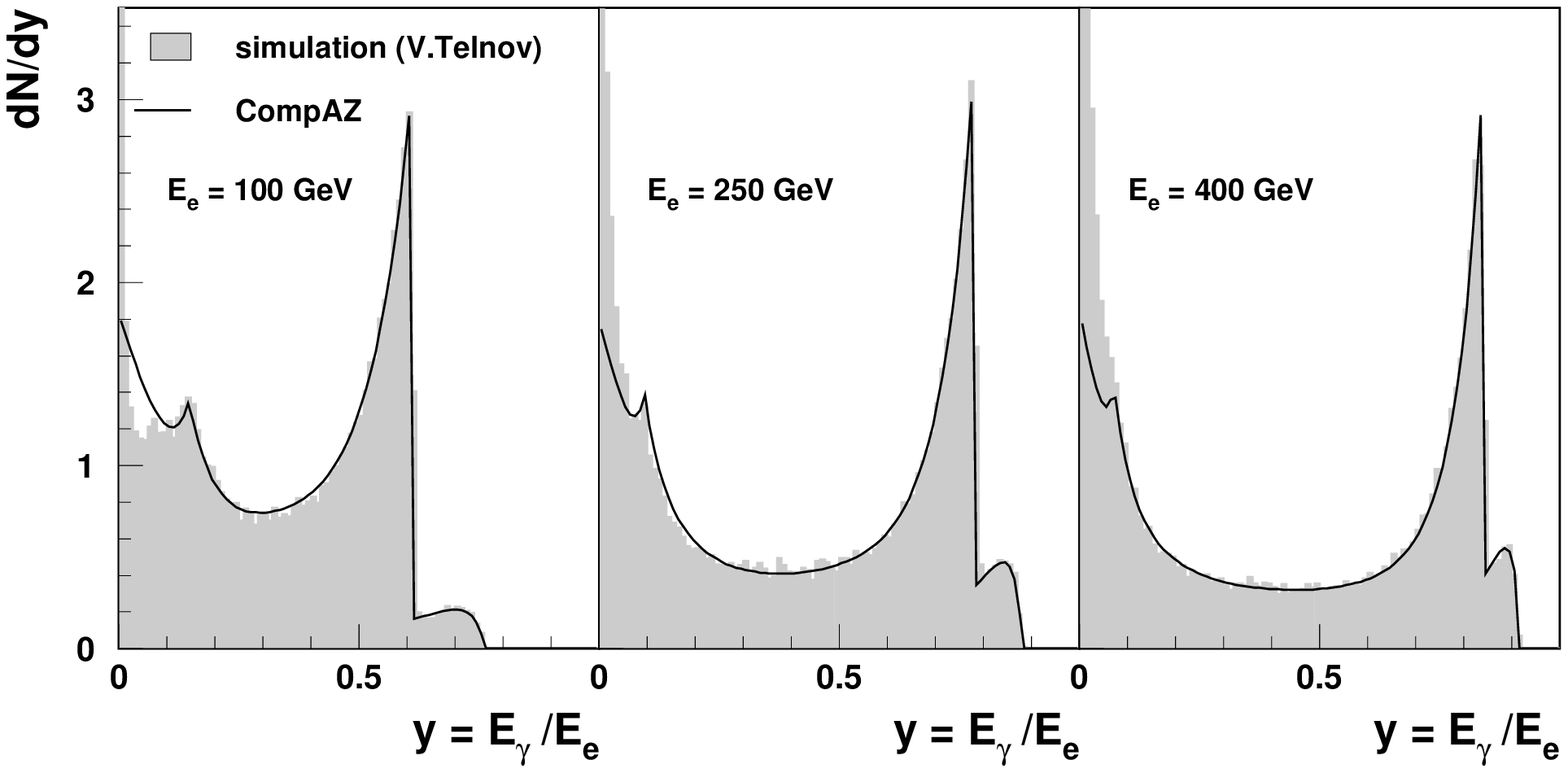,height=\figheight,clip=}
  \end{center}
  \caption{Comparison of the  photon-energy distribution 
        from the CompAZ parametrization with the distribution
        obtained from a full simulation of luminosity spectra 
        \cite{TEL95,TEL01},
        for three electron-beam energies, as indicated in the plot.
        The cuts imposed  on the energy of the second photon are  
        40, 150 and 260 GeV, respectively.
           }
  \label{fig:compaz_e}
\end{figure}

\begin{figure}[tbp]
  \begin{center}
  \epsfig{figure=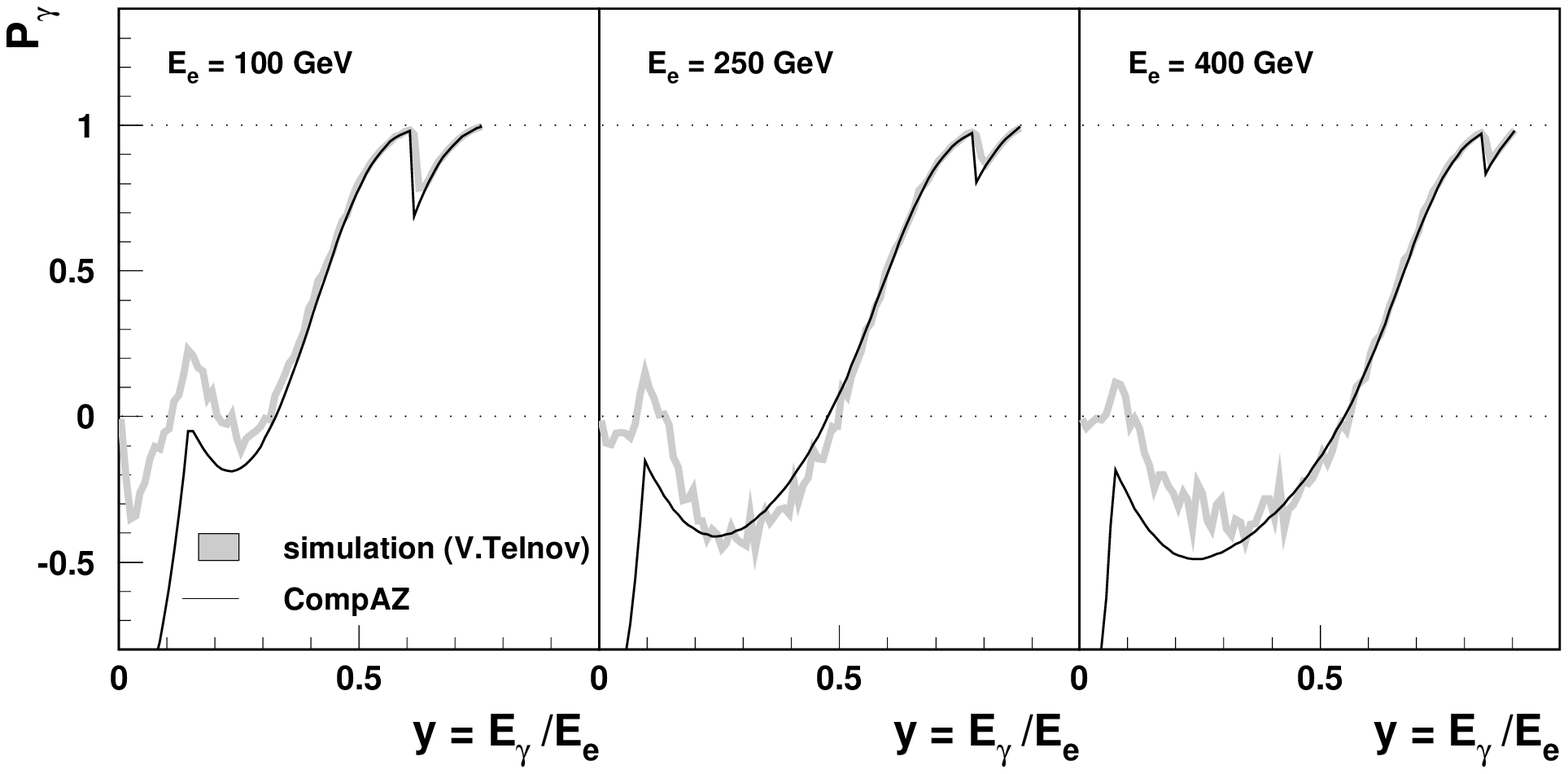,height=\figheight,clip=}
  \end{center}
  \caption{Comparison of the  photon polarization 
        from the CompAZ parametrization with the distribution
        obtained from a full simulation of luminosity spectra \cite{TEL95,TEL01},
        for three electron-beam energies, as indicated in the plot.
        The cuts imposed on the energy of the second photon are 
        40, 150 and 260 GeV, respectively.
           }
  \label{fig:compaz_p}
\end{figure}

In the present analysis, the  CompAZ parametrization \cite{compaz} of
the photon collider luminosity spectra at TESLA  is used.
The program is based on a realistic simulation of \gg\ 
spectra \cite{TEL95,TEL01}, which became available recently.
Figures \ref{fig:compaz_e} and \ref{fig:compaz_p} show
the comparison of the  photon energy and polarization 
distribution from CompAZ with the results of a full simulation.
The high-energy part of the spectrum, relevant to  the
present analysis, is very well described.
%

Results presented in this paper assume one year of photon collider
running at nominal luminosity.
The integrated luminosity that can be obtained for \gg\ collisions 
increases from about 600$\;\rm fb^{-1}$ for running at $\sqrt{s_{ee}}= 305$ GeV
to about  1000$\;\rm fb^{-1}$ for  $\sqrt{s_{ee}}= 1 $ TeV \cite{tdr_pc}.
This corresponds to an integrated luminosity in the high-energy
part of the spectra ($W_{\gamma \gamma} > 0.8 \; W_{max} $) 
of 75 to 115$\;\rm fb^{-1}$.

\subsection{Simulation}

The event generation was done with PYTHIA~6.152~\cite{PYTHIA}.
For $\gamma \gamma \rightarrow W^+ W^-$,
events generated with PYTHIA were reweighted to take 
into account photon polarization as well as Higgs production 
and interference. 
For  $\gamma \gamma \rightarrow \z0  \z0 $,
events were generated according to the cross-section
formula for direct $\z0  \z0 $ production \cite{cros_zz}
and reweighted for Higgs  contribution.

The fast TESLA detector simulation program SIMDET version 3.01 \cite{SIMDET}
was used to model detector performance.

\begin{figure}[tbp]
  \begin{center}
  \epsfig{figure=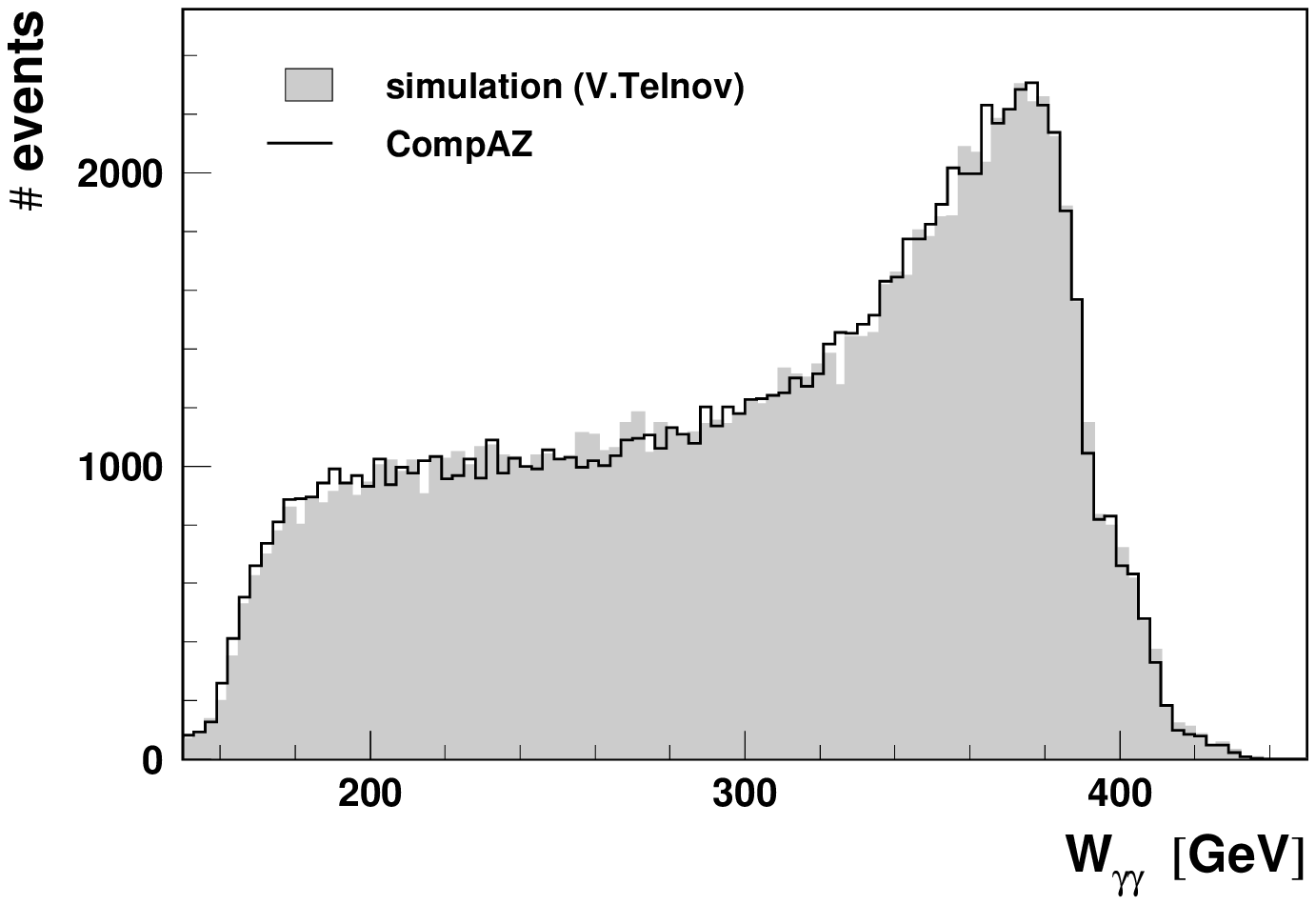,height=\figheight,clip=}
  \end{center}
  \caption{Distribution of the $\gamma \gamma$ center-of-mass energy 
         for $\gamma \gamma \rightarrow W^+ W^-$ events 
        generated with PYTHIA, for an electron beam energy of 250 GeV.
        The sample of events generated using
        the CompAZ parametrization is compared with the sample
        generated with the luminosity spectra from a full 
        simulation \cite{TEL95,TEL01}.
        The generated events were reweighted for photon polarizations.
          }
  \label{fig:compaz_wgg}
\end{figure}

\subsection{Event selection}

\begin{figure}[tbp]
  \begin{center}
  \epsfig{figure=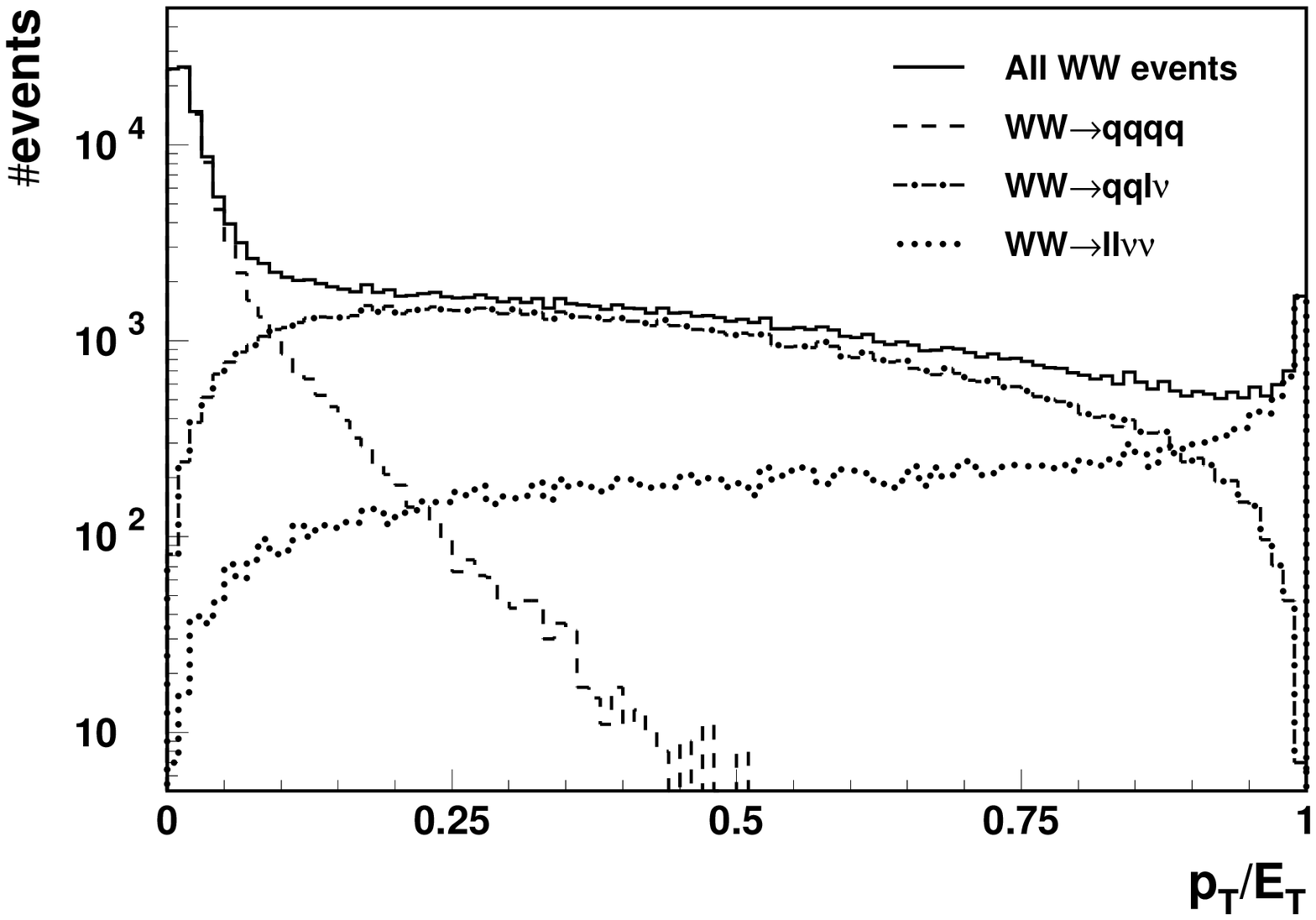,height=\figheight,clip=}
  \end{center}
  \caption{ Measured ratio of the total transverse momentum $p_T$
         to the total transverse energy $E_T$ for 
         $\gamma \gamma \rightarrow W^+ W^-$ events
         simulated with PYTHIA and SIMDET, for
         a primary electron-beam energy of 250~GeV.
         Contributions from different $W^\pm$ decay channels
         are indicated.
          }
  \label{fig:ptcut_ww}
\end{figure}

\begin{figure}[tbp]
  \begin{center}
  \epsfig{figure=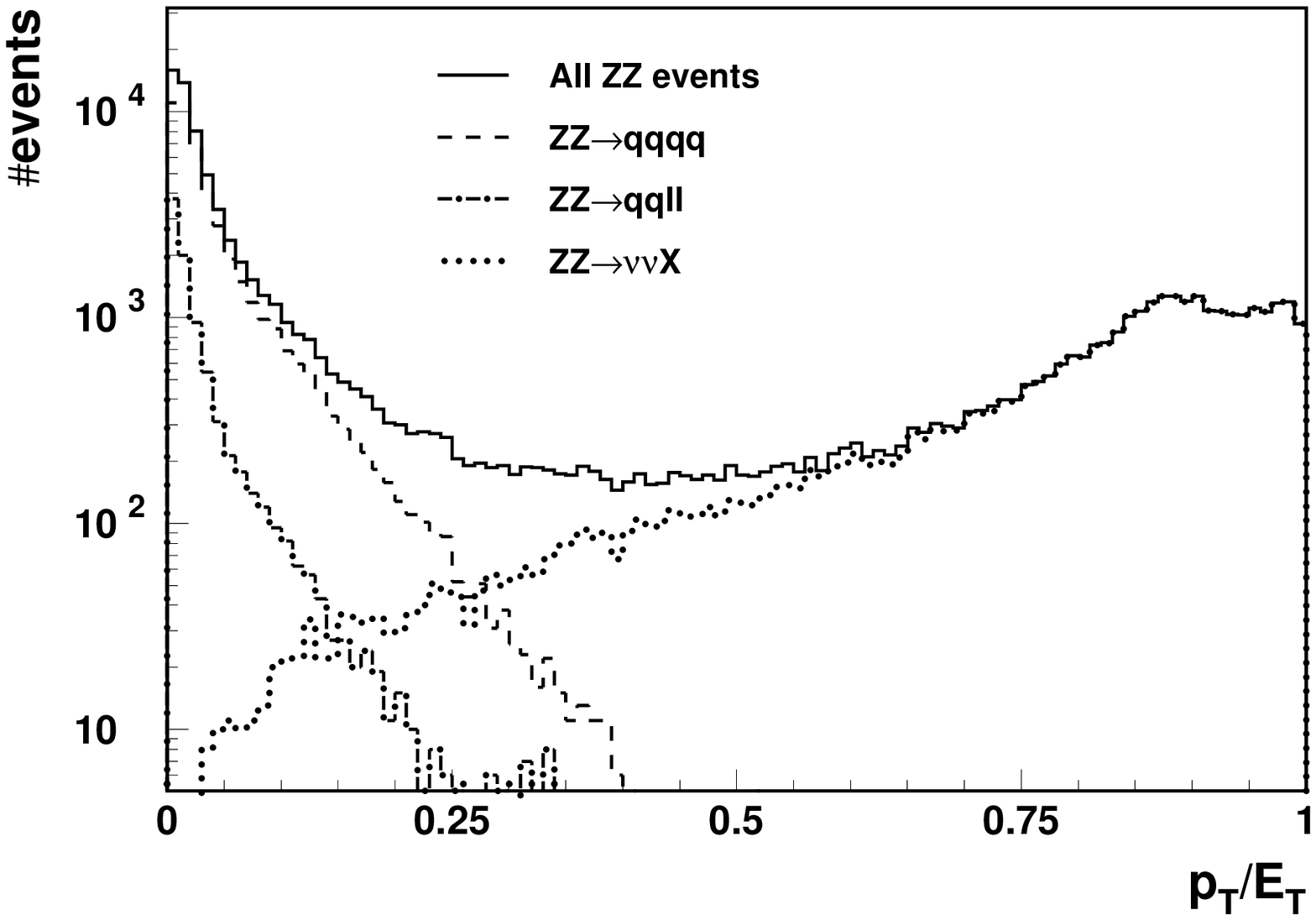,height=\figheight,clip=}
  \end{center}
  \caption{ As in Fig.~\ref{fig:ptcut_ww}, 
           for $\gamma \gamma \rightarrow \z0  \z0 $ events.
          }
  \label{fig:ptcut_zz}
\end{figure}

The selection cuts were applied to select $W^+ W^- \rightarrow qq\bar{q}\bar{q}$
and $\z0  \z0  \rightarrow l\bar{l}q\bar{q}$ events.
The first cut is based on the requirement that four
jets can be reconstructed with the Durham algorithm, with $y_{cut} \ge 0.0004$ 
(from the point of view of the algorithm, a jet can also be a single particle).
To remove the background from semileptonic $W^\pm$ decays and from 
$\z0 $ decays to neutrinos, a cut on the ratio of the total 
transverse momentum $p_T$ to the total transverse energy $E_T$ 
was applied.
Distributions of the measured ratio $p_T / E_T$ for
$\gamma \gamma \rightarrow W^+ W^-$  and
$\gamma \gamma \rightarrow \z0  \z0 $ events
are shown in Figs. \ref{fig:ptcut_ww} and \ref{fig:ptcut_zz}.
Neutrinoless $W^\pm$ and $\z0$ decays result in a narrow peak at
$p_T / E_T \sim 0$.
A cut  $p_T / E_T \; < \; 0.1$  was applied for
both $W^+ W^- $ and $\z0  \z0  $  events.
%
%

\subsubsection{$W^+ W^-$ events}

\begin{figure}[tbp]
  \begin{center}
  \epsfig{figure=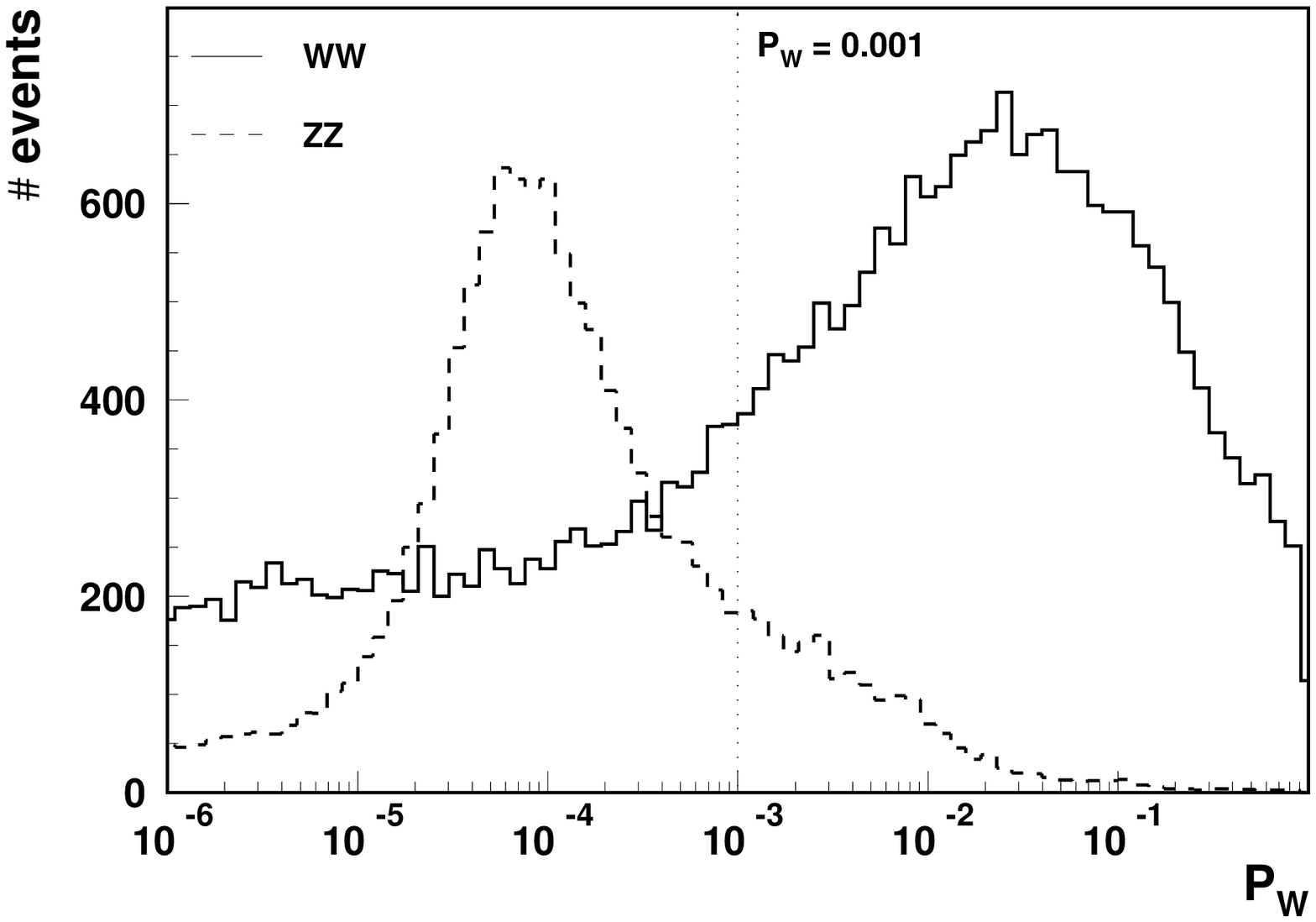,height=\figheight,clip=}
  \end{center}
  \caption{Distribution of the probability $P_W$, of four jets
          coming from  $W^+ W^-$ pair production
         (for the jet configuration with the highest  $P_W$ value),
          for $W^+ W^-$ and $\z0 \z0$ events directly produced
          in \gg\ collisions, for primary electron-beam energy of 250 GeV.
          Only hadronic $W^\pm$ and $\z0$ decays are considered.
          The dotted line indicates the cut used in  $W^+ W^-$ event
           selection.
          }
  \label{fig:pwcut}
\end{figure}

For  $W^+ W^- \rightarrow qq\bar{q}\bar{q}$ events,
each of the four reconstructed jets should
be recognized as a ``true'' hadronic jet.
The simple algorithm used for removing jets consisting of too few particles 
(usually isolated leptons) requires that the number of particles in a jet 
 be greater than 5, or that the fraction of energy carried by the most 
energetic particle  be smaller than 95\%.
The next step is to reconstruct two $W^\pm$ bosons.
There are three possible ways of selecting two pairs out of four jets.
The configuration with the highest probability  is taken,
 defined as
\begin{eqnarray*}
P_W & = & \prod_{W_1,W_2} \frac{M_W^2 \Gamma_W^2}{
          \left( m_{jj}^2 - M_W^2 \right)^2 + M_W^2 \Gamma_W^2} \; .
\end{eqnarray*}
The product runs over two jet pairs, $m_{jj}$ is the jet-pair
invariant mass, $M_W$ and $\Gamma_W$ are the mass and the total width
of the $W$ boson.
Distribution of the probability $P_W$ for $W^+ W^-$ and $\z0 \z0$ 
events from non-resonant production is shown in Fig.~\ref{fig:pwcut},
for a primary-electron beam energy of 250~GeV.
Only hadronic $W^\pm$ and $\z0$ decays are considered.
%
For most $W^+ W^-$ events the   $P_W$ value is above $10^{-3}$ 
whereas $\z0 \z0$ events result in smaller values.
Separation is clear, although the mass resolution (see section \ref{sec:resol})
is comparable with the mass difference between 
$W^\pm$ and $\z0$.
For other backgrounds 
(e.g. $\gamma \gamma \rightarrow q \bar{q} g g $),
much smaller values of $P_W$ are expected.
For the final selection of $W^+ W^-$ events, 
$P_W >0.001$ was required.

\begin{figure}[tbp]
  \begin{center}
  \epsfig{figure=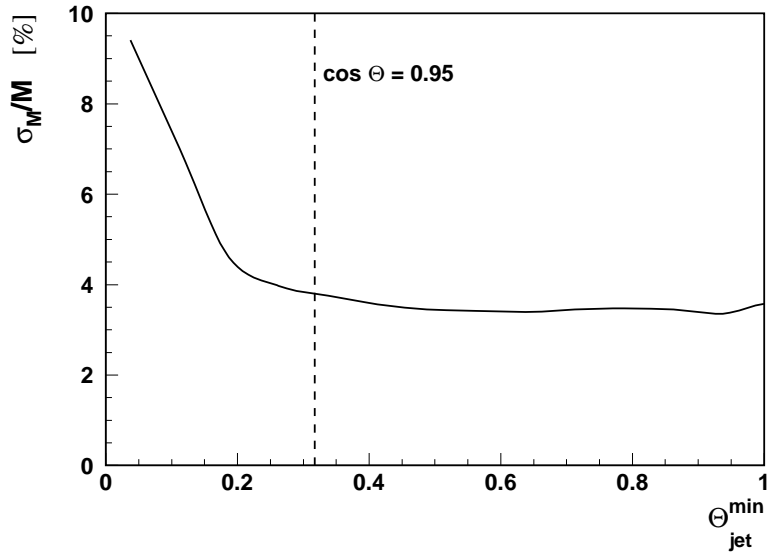,height=\figheight,clip=}
  \end{center}
  \caption{Relative resolution in the reconstructed four-jet
           invariant mass as a function of the minimum jet
           scattering angle $\theta^{min}_{jet}$, for
           $\gamma \gamma \rightarrow W^+ W^-$ events.
           The simulation was done using PYTHIA and SIMDET, for
           a primary electron-beam energy of 250 GeV.
           The dashed line indicates the angle corresponding
           to  $ \cos \theta=0.95$.
          }
  \label{fig:cjcut}
\end{figure}

For the reconstruction of $h \rightarrow W^+ W^-$ events, a 
good mass resolution is essential.
Figure \ref{fig:cjcut} shows the relative resolution 
in the reconstructed four-jet invariant mass 
as a function of the minimum jet scattering angle $\theta^{min}_{jet}$.
The mass resolution deteriorates significantly if any of the
jets is emitted at an angle smaller than about 0.3 with respect
to the beam axis.
To preserve good mass resolution, the cut on the jet angle
$|\cos \theta_{jet}| < 0.95$ is imposed for all jets.
%
After all cuts, the efficiency for selecting 
$\gamma \gamma \rightarrow W^+ W^-$ events is between 20\% for
\wgg \,$\sim$ ~200 GeV  and 16\% for \wgg \,$\sim$ ~400 GeV.
\footnote{The efficiencies include 47\% probability for 
both $W^\pm$ to decay into hadrons.}

\subsubsection{$\z0  \z0$  events}

For  $\z0 \z0 \rightarrow l\bar{l}q\bar{q}$ events,
two out of the four reconstructed jets should
be recognized as  hadronic jets and the other two
as isolated leptons.
Only $\z0$ decays into $e^+ e^-$ and $\mu^+ \mu^-$ are considered.
Invariant masses of the lepton pair and of the two jets 
are required to be close to the $\z0$ mass.
The possible background is rejected by the cut on the probability $P_Z$
\begin{eqnarray*}
P_Z & = & \frac{M_Z^2 \Gamma_Z^2}
               {\left( m_{jj}^2 - M_Z^2 \right)^2 + M_Z^2 \Gamma_Z^2}
   \cdot  \frac{M_Z^2 \Gamma_Z^2}
               {\left( m_{ll}^2 - M_Z^2 \right)^2 + M_Z^2 \Gamma_Z^2},
\end{eqnarray*}
where $m_{ll}$ is the lepton-pair
invariant mass, $M_Z$ and $\Gamma_Z$ are the mass and the total width
of the $\z0$ boson.
For the final selection of $\z0  \z0 $ events 
$P_Z >0.001$ was required.
After all cuts, the selection efficiency for $\z0  \z0 $ events is only about 5\%,
mainly due to the small branching ratio for the considered channel
(9.4\% for  $\z0 \z0 \rightarrow l\bar{l}q\bar{q}, \; l=e,\mu$). 
However, the final sample is very clean.
No events from the large simulated samples of $W^+ W^-$ and $f\bar{f}$
production remain after $\z0  \z0$ selection cuts.
 
\subsection{Mass resolution}

\label{sec:resol}

\begin{figure}[tbp]
  \begin{center}
  \epsfig{figure=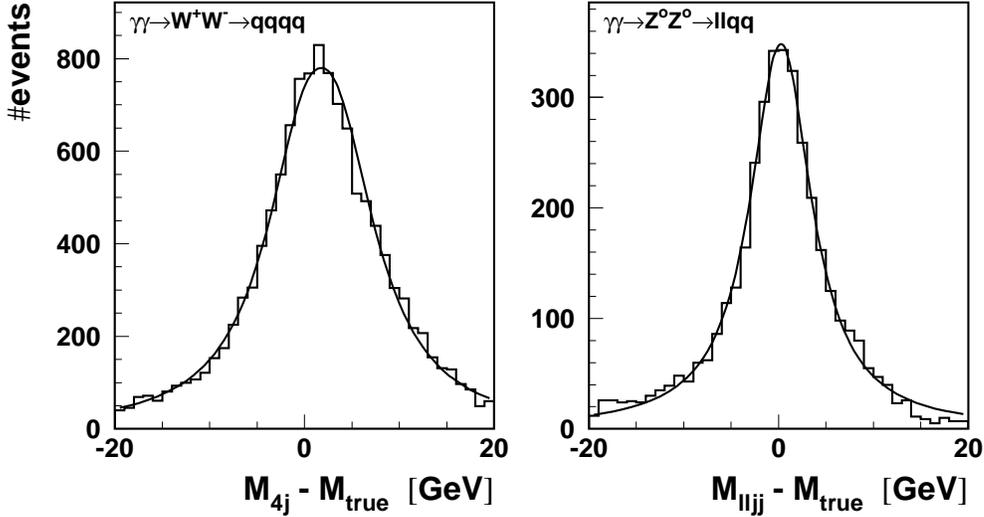,height=\figheight,clip=}
  \end{center}
  \vspace{-1cm}
  \caption{Resolution in the reconstructed $\gamma \gamma$
           invariant mass for the selected
           $\gamma \gamma \rightarrow W^+ W^-$ events (left plot)
           and $\gamma \gamma \rightarrow \z0  \z0 $ events
           (right plot), for 200 GeV $< \; W_{\gamma \gamma} \; <$ ~300 GeV.
           Events were simulated with the PYTHIA and SIMDET programs, for
           a primary electron-beam energy of 250 GeV.
           The curves correspond to the parametrization used to
           describe mass resolution (see text for more details).
          }
  \label{fig:mres_wwzz}
\end{figure}

 Figure \ref{fig:mres_wwzz} shows  the error on 
the reconstructed $\gamma \gamma$ invariant mass 
for the selected  $\gamma \gamma \rightarrow W^+ W^-$ 
 and $\gamma \gamma \rightarrow \z0  \z0 $ events,
for 200 GeV $< \; W_{\gamma \gamma} \; <$ 300 GeV.
The distributions of the difference between the reconstructed and
the true invariant mass, as given by the detector simulation,
have a long tail. 
It cannot be described by a Gaussian resolution;
however, a good description of the detector performance is given by
the modified Breit--Wigner curve:
\begin{eqnarray*}
P(m) & \sim & 
\frac{\Gamma^\alpha}{4 (m - m_{true})^\alpha + \Gamma^\alpha}.
\end{eqnarray*}
The parameters $\Gamma$ and $\alpha$ were fitted to the
simulation results in small mass bins.
The value of  $\Gamma$ 
for a four-jet invariant-mass reconstruction changes from
about 6.5 GeV at \wgg \, = ~200 GeV to about 13 GeV at \wgg \,= ~400 GeV.
For the $l\bar{l} q \bar{q}$ final state, $\Gamma$ changes from
about 5.5 GeV at \wgg \, = ~200 GeV to about 7.5 GeV at \wgg \, = ~400 GeV.
The value of $\alpha$ is between 2 and 2.5 for both samples.
%

\subsection{Parametrization}

The invariant-mass resolution obtained from a  full simulation of $W^+ W^-$  
and $\z0 \z0$ events based on the PYTHIA and SIMDET programs, 
has been parametrized
as a function of the \gg\ centre-of-mass energy \wgg .
By convoluting  this parametrization with the CompAZ photon energy spectra
and the cross sections for $\gamma \gamma \rightarrow W^+ W^-$ 
and $\gamma \gamma \rightarrow \z0 \z0$ processes, a
detailed description of the expected mass spectra can be obtained 
without time-consuming event generation.
%
\begin{figure}[tbp]
  \begin{center}
  \epsfig{figure=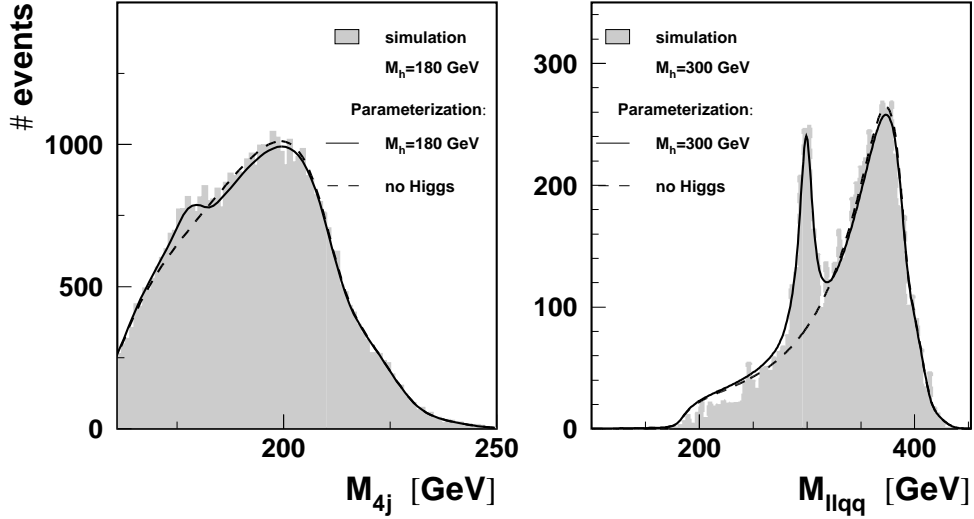,height=\figheight,clip=}
  \end{center}
  \vspace{-1cm}
  \caption{Distribution of the reconstructed invariant mass
        for $\gamma \gamma \rightarrow W^+ W^-$ events with a 
        SM Higgs-boson mass of 180 GeV and an 
        electron-beam energy of 152.5 GeV (left plot) 
        and
        for $\gamma \gamma \rightarrow \z0  \z0 $ events, with a 
        SM Higgs-boson mass of 300 GeV and an 
        electron-beam energy of 250 GeV (right plot). 
        Results from the simulation  based  on PYTHIA 
        and on the SIMDET detector simulation (histogram) are 
        compared with the distribution
        obtained  by the numerical convolution of the cross-section 
        formula with the CompAZ photon energy spectra
        and parametrization of the detector resolution (solid line).
        The distribution expected without the Higgs contribution is
        also shown (dashed line).
          }
  \label{fig:convol}
\end{figure}
 In Fig. \ref{fig:convol} we show  the distribution 
of the reconstructed invariant mass
for $\gamma \gamma \rightarrow W^+ W^-$ events (with
an SM Higgs-boson mass of 180 GeV 
and an electron beam energy of 152.5 GeV) 
 and  for $\gamma \gamma \rightarrow \z0  \z0 $ events (with
a SM Higgs-boson mass of 300 GeV 
and an  electron beam energy of 250 GeV).
 Results coming from the full event simulation  based  on PYTHIA
 and SIMDET (events reweighted for the Higgs signal) 
 are  compared with the distribution
 obtained  by the numerical convolution of the cross-section 
 formula (including the Higgs signal) with the CompAZ spectra
 and detector resolution.
A very good description of the measured mass distribution is obtained.
Using this approach, the expected mass distributions were calculated 
for different Higgs-boson masses (from 170 to 390 GeV) and for
different centre-of-mass energies of colliding electron beams
(305, 362, 418 and 500 GeV).


\section{Results}
\label{sec:results}

\begin{figure}[tbp]
  \begin{center}
  \epsfig{figure=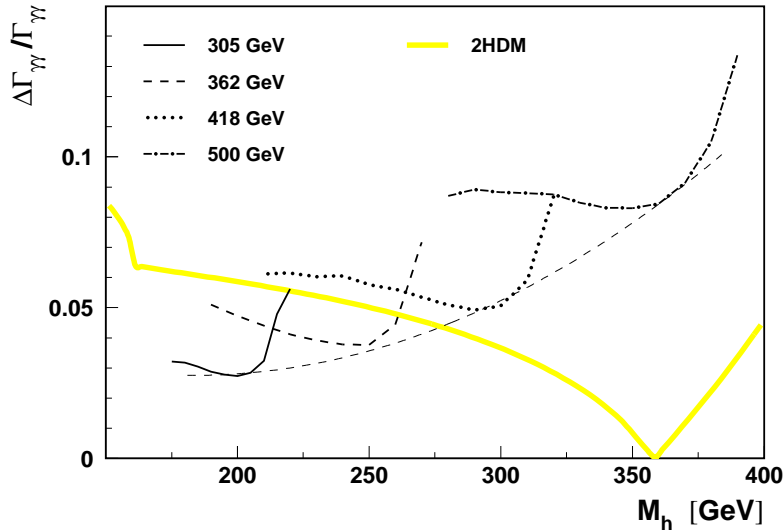,height=\figheight,clip=}
  \end{center}
  \vspace{-1cm}
  \caption{Average statistical error in the determination of the
          Higgs-boson width $\Gamma_{\gamma \gamma}$, expected
          after one year of photon collider running, from the simultaneous fit           to the observed $W^+ W^-$ and $ZZ$ mass spectra,
          as a function
          of the Higgs-boson mass $M_h$.
          Results are given for various centre-of-mass
          energies of colliding electron beams $\sqrt{s_{ee}}$, as indicated
          in the plot. The yellow (tick light) band shows the size of 
          the deviation
          expected in the SM-like  2HDM~(II) 
          with an additional contribution due to the charged
          Higgs-boson of  mass $M_{H^+}=800$ GeV. 
          The thin dashed line is included to guide the eye. 
          }
  \label{fig:final_1}
\end{figure}

Based on the parametric description of the expected mass distributions, 
a number of  experiments were simulated, each corresponding to one year
of TESLA photon collider running at nominal luminosity,
by generating Poisson-distributed
numbers of observed events in bins of reconstructed mass.
This was done
for different Higgs-boson masses and different electron beam energies.
The ``theoretical'' distributions were then fitted, simultaneously to
the observed $W^+ W^-$ and $\z0 \z0$ mass spectra, with the 
width \ggg\ considered the only free parameter.
Figure \ref{fig:final_1} shows the average statistical 
error on the fitted \ggg\ value expected
after one year of photon collider running, for different
masses of the Higgs-boson  and different centre-of-mass
energies of colliding electron beams.
Standard Model Higgs-boson branching ratios are assumed.
For the beam energy range considered in this analysis
the two-photon width of the Higgs-boson can be measured
with an accuracy of 3 to 8\%, for Higgs-boson masses between
200 and 350 GeV.  
The precision of the measurement deteriorates with the Higgs-boson mass.
For masses above about 275 ~GeV the expected statistical error
is larger than an effect expected in the SM-like 
Two Higgs Doublet Model (2HDM~(II))
with charged Higgs-boson mass $M_{H^+}=800$ GeV.

\begin{figure}[tbp]
  \begin{center}
  \epsfig{figure=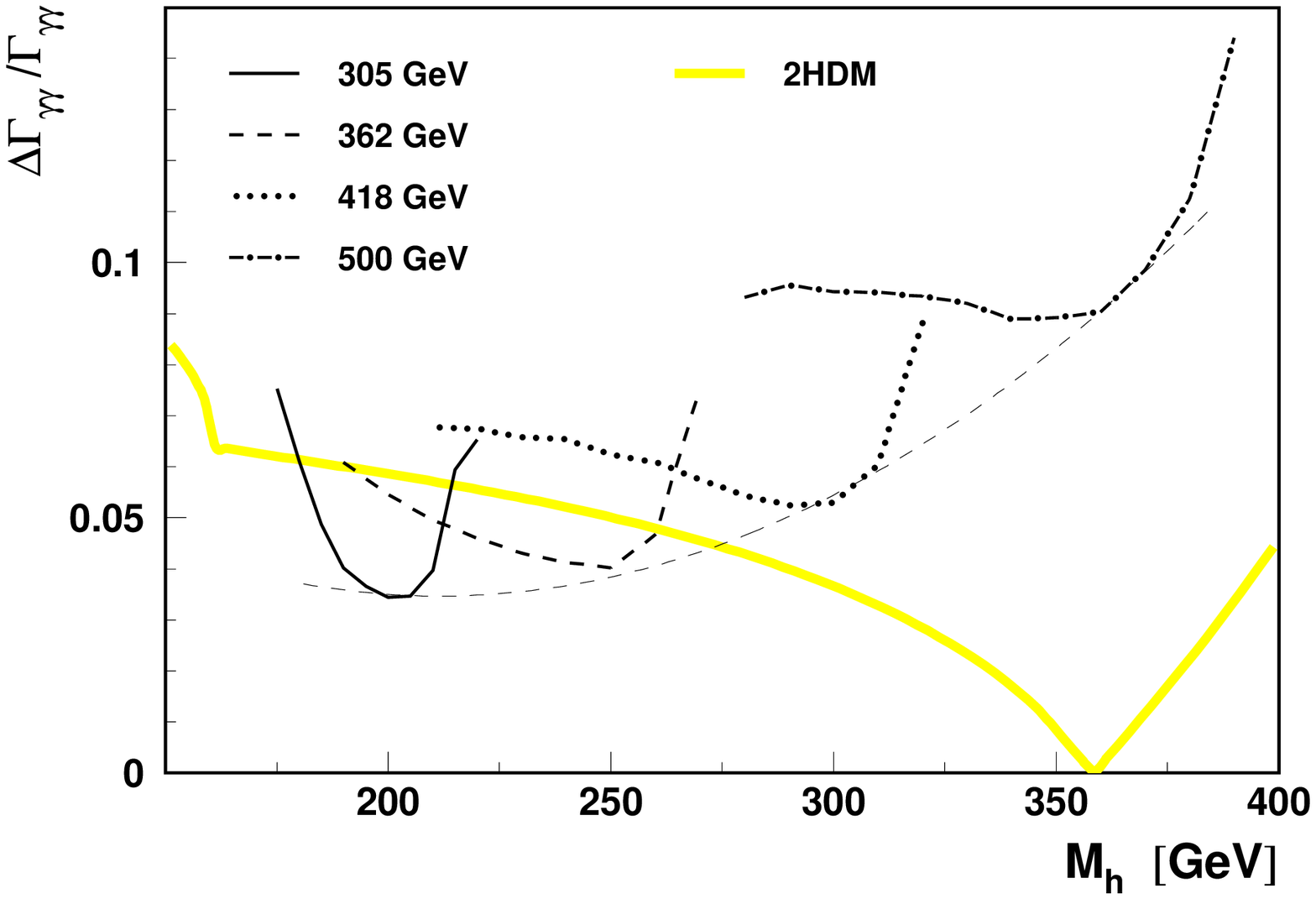,height=\figheight,clip=}
  \epsfig{figure=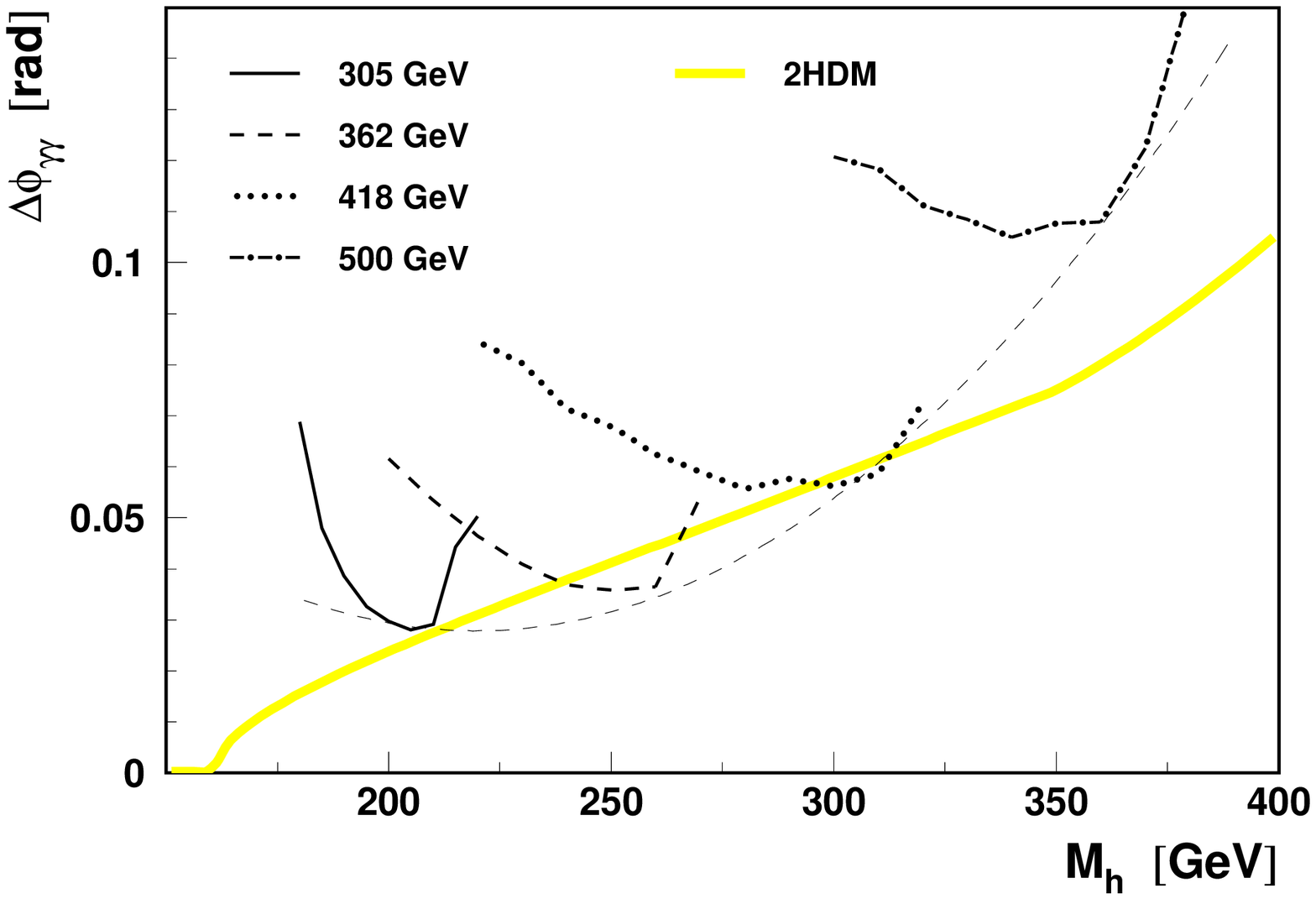,height=\figheight,clip=}
  \end{center}
  \caption{As in Fig.~\ref{fig:final_1}, now for a simultaneous 
           determination of the  Higgs-boson width \ggg\  and
           phase  \pgg\,
           results for the width and the phase are shown in upper and lower
           plot, respectively.
%
          }
  \label{fig:final_5}
\end{figure}

Figure \ref{fig:final_5} shows the average statistical 
error expected from the {\bf simultaneous fit} of \ggg\ 
and the $h\rightarrow \gamma \gamma$ amplitude phase, 
\pgg\ (the two-parameter fit)  to the observed WW and ZZ mass spectra.
The error on the extracted two-photon width of the Higgs-boson
increases only slightly.
On the other hand, 
for Higgs-boson masses between 200 and 350 GeV  
the phase of the amplitude can be measured
with a statistical accuracy between 30 and 100 mrad.  
%
For masses between 210 and 310 GeV  
the expected statistical error
is lower than an effect (i.e. departure from SM predictions)
expected in the SM-like  2HDM~(II)
already after one year of photon collider running.  
This demonstrates that the phase measurement opens a new window to 
searches of processes beyond the Standard Model, in the Higgs-boson
mass range where the width measurement is little sensitive to new physics.
%


\section{Summary}
\label{sec:concl}

Production of the Standard Model Higgs-boson at the photon collider at
TESLA has been studied for the Higgs-boson masses above 150 GeV. 
In the considered mass range, large 
interference effects are expected in the $W^+ W^-$ decay channel. 
By  reconstructing  $W^+ W^-$ and $\z0  \z0  $ final states, 
not only the $h \rightarrow \gamma \gamma$
partial width \ggg\ can be measured, but also the phase of the scattering amplitude. 
For Higgs-boson masses around 350 GeV, the amplitude phase \pgg\
is expected to be more sensitive to the loop contributions of new,
heavy charged particles than  the \ggg\ itself.

The analysis presented here  was based on the full event simulation,
including the realistic luminosity spectra and simulation of 
detector effects. 
The parametrization of measured mass distributions was made possible
by using the CompAZ spectra parametrization.
Results of the fits performed for different Higgs-boson masses
and at different electron beam energies indicate, that
with a proper choice of the beam energy, the $\gamma \gamma$
partial width can be measured with an accuracy of 3 to 8\%
and the  phase of the amplitude with an accuracy between 30 and 100~mrad.
This opens a new window to  the precise
determination of the Higgs-boson couplings
and therefore to  searches of ``new physics''.

%
%

%

\subsection*{Acknowledgments}

We would like to thank  our colleagues from the 
Extended Joint ECFA/DESY Study on 
Physics and Detectors for a Linear Electron-Positron Collider
for many useful discussions, comments and suggestions.

The work was partially sponsored
by BMBF-KBN collaboration program TESLA. M.K.~acknowledges partial
support by Polish Committee for Scientific Research, Grant 
2~P~03~B~05119 (2002), 5 P03B 121 20 (2002), and by the European Community's
Human Potential Programme under contract HPRN-CT-2000-00149 Physics
at Colliders.
M.K. and A.F.\.Z would like to thank DESY Directoriate for kind hospitality 
during their stay at DESY, where part of the work was completed.

\clearpage



\end{document}